\documentclass[%
 reprint, 
superscriptaddress,
 amsmath,amssymb,
 aps,
pra,
]{revtex4-2}
\usepackage{comment}
\usepackage{graphicx}
\usepackage{dcolumn}
\usepackage{bm}
\usepackage{braket}

\begin{document}

\preprint{APS/123-QED}

\title{Indistinguishablity from dephased emitters\\using combined plasmonic-dielectric cavities}%
\author{Anastasios Fasoulakis}
\affiliation{Institute for Physics and Astronomie, Technical University Berlin, Hardenbergstraße 36, 10623 Berlin,
Germany}
\author{Ross C. Schofield}
\affiliation{Blackett Laboratory, Imperial College London, London, SW7 2AZ, UK
}
\author{Rupert F. Oulton}
\affiliation{Blackett Laboratory, Imperial College London, London, SW7 2AZ, UK
}
\author{Alex S. Clark}
\affiliation{Quantum Engineering Technology Labs, H. H. Wills Physics Laboratory and School of Electrical, Electronic, and Mechanical Engineering, University of Bristol, BS8 1FD, UK}
%
\date{\today}

\begin{abstract}
The concept of cavity funneling has emerged recently as a promising route towards creating indistinguishable photons from highly dephased emitters. So far, all suggested solutions are solely based on dielectric cavities that require extremely high quality factors that are difficult to reach at visible wavelengths. Here we suggest a hybrid funneling architecture where a dephased emitter is coupled to a plasmonic nanoresonator that is enclosed by an outer dielectric cavity. The estimated lower limit of the outer cavity quality factor is found to be $\sim2$ orders of magnitude lower compared to a cascaded cavity system. Furthermore, the surrounding topology of our approach allows for a partial direct coupling between the emitter and the outer cavity which in turn can increase the overall system extraction efficiency $\left(\beta\right)$ by a factor of 12, boosting the probability of photon collection.
\end{abstract}

\maketitle


Indistinguishability between photons is a core requirement of any optical quantum information application, including linear optical quantum computing \cite{RevModPhys.79.135,maring2024versatile}, quantum simulations \cite{aspuru2012photonic,wang2020integrated,sparrow2018simulating} and quantum key distribution (QKD) \cite{Noblet:23,zhan2025experimental,Ge:24}. However, all solid state emitters experience strong pure dephasing at elevated temperatures. This significantly broadens the spectral width of the emitted photons by orders of magnitude compared to their Fourier-limited linewidth, thus rendering them unsuitable for quantum information processing. Mitigating this by shortening the emitters' lifetime via the Purcell effect is practically unreachable with the required enhancement being well beyond the current state of the art values \cite{Gritsch:23,doi:10.1021/acsphotonics.4c01873,doi:10.1021/acs.nanolett.4c06405}. 

An alternative approach presented by Grange \textit{et al.} proved that if the decay rate $\kappa$ of a dielectric cavity is sufficiently low, then weakly coupling it with a dephased emitter under an equally low coupling rate $g$ effectively `funnels' the broad linewidth photons into the cavity's narrow mode \cite{PhysRevLett.114.193601}. As a result, the photons that exit the cavity should present high levels of indistinguishability $\left(I\right)$ along with a photon extraction efficiency $\left(\beta\right)$ that is orders of magnitude higher than that of a simple linear filter. However, in such a scheme the relevant quality factor would need to be extremely high, up to values of $Q\sim10^7$, depending on the emitter lifetime and emission wavelength. Even cascading two dielectric cavities instead of a single one can only reduce this requirement by two orders of magnitude \cite{PhysRevLett.122.183602}. To the best of our knowledge, at visible wavelengths where room temperature emitters mainly operate $\left(\lambda<700~\text{nm}\right)$, the highest measured $Q$-factors that have been achieved in cavity structures only extend up to the order of $Q\sim10^4$ \cite{10.1063/1.4904909,10.1063/1.4992118,10.1063/1.5120120,ding2024high}. Thus, the practical infeasibility of this approach still persists.

Combining dielectric resonators with plasmonic nanostructures has recently been shown to improve the performance of various optical systems by combining the benefits of both components in a hybrid architecture\cite{doi:10.1021/acsphotonics.7b00953,PhysRevA.111.043507,YAN2024130923}. In this work we theoretically investigate the implementation of cavity funneling via coupling a highly dissipative emitter to a combined structure consisting of a plasmonic nanoresonator embedded inside a typical Fabry-Perot (F-P) dielectric cavity. Even though the emitter-plasmonic resonator system operates in the weak coupling regime, eventually it behaves like an equivalent effective emitter of a much larger decay rate due to the tiny mode volume that is offered by the metallic nanostructure and its very broad spectrum. This large decay rate then enables funneling operations of up to $\sim2$ orders of magnitude higher values of the outer cavity's decay rate compared to those predicted in the cascaded cavity system \cite{PhysRevLett.122.183602}. Moreover, in principle, the nested geometry of our approach uniquely provides an additional direct photon exchange channel between the emitter and the outer cavity. As illustrated here, exploiting such a channel can further enhance the final photon extraction efficiency by an additional factor of 12, hence mitigating the effect of the additional losses that occur due to the presence of metallic surfaces.  
\begin{figure}[b]
\centering
\includegraphics[width=6.5cm]{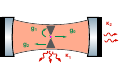}
\caption{\label{fig:setup} A schematic of the suggested system. A single emitter is coupled to a plasmonic structure, such as a bowtie, and they both are placed inside an outer dielectric F-P cavity.}
\end{figure}

The system of interest is presented in Fig.~\ref{fig:setup}. We consider an arbitrary quantum emitter whose free space excited state decay rate is hereby depicted as $\Gamma_1$ and whose pure dephasing rate is expressed as $2\gamma^*$. The decay rates of the plasmonic device and the dielectric cavity are set as $\kappa_1$ and $\kappa_2$ correspondingly. The dipolar coupling rate between the emitter and the plasmonic device is set as $g_1$, while that between the plasmonic particles and the outer cavity is $g_2$. Since large values of both $g_1$ and $\kappa_1$ are essential in our system, a bowtie antenna is used for the plasmonic resonating device, due to their typically small mode volumes and low quality factors. However, the presented theory is independent of the type of plasmonic antenna and F-P cavity chosen. The findings remain the same for any combination of resonators provided that the $g_1$, $g_2$, $\kappa_1$ and $\kappa_2$ are achieveable. The suggested geometry also allows for an additional direct coupling between the emitter and the outer cavity, which we label as $g_0$. 

Assuming that only a single mode of the outer cavity's resonant field can overlap with the emitter spectrum, the indistinguishability of exiting photons is given by the following expression,
\begin{equation}
I = \frac{ \int_{0}^\infty \int_{0}^\infty |\langle b^\dagger\left(t+\tau\right) b\left(t\right)\rangle|^2 d\tau dt}{\int_{0}^\infty \int_{0}^\infty \langle b^\dagger\left(t\right) b\left(t\right)\rangle\langle b^\dagger\left(t+\tau\right) b\left(t+\tau\right)\rangle d\tau dt} \;,
\label{eq:I_definition}
\end{equation}
where $b$ and $b^\dagger$ are the bosonic creation and annihilation operators of that mode \cite{PhysRevLett.114.193601,PhysRevLett.59.2044,PhysRevB.87.081308}. Since our system is studied per single excitation cycle, there cannot be more than a single photon present at any time, hence the relevant photon extraction efficiency can be defined as 
\begin{equation}
    \beta = \kappa_2 \int_{0}^\infty \langle b^\dagger\left(t\right) b\left(t\right) \rangle dt \,.
\label{eq:beta_definition}
\end{equation}
This metric can be interpreted as the probability per excitation cycle that a single photon from the resonant mode of the cavity would escape into a measurement setup. Even though $\beta$ is relatively low in funneling systems, the achievable values of the product $\beta\cdot I$ are orders of magnitude higher than those that can be obtained by similar linear filters \cite{PhysRevLett.114.193601}. Usually, this advantage can be gauged quantitatively via the funneling ratio $F$ which can be expressed in logarithmic form as 
\begin{equation}
    F = 10 \log\left(\frac{2\gamma^*}{\Gamma_1}\beta I\right)
    \label{eq:funnel_ratio_definition} \,.
\end{equation}

\begin{figure}[t]
\centering
\includegraphics[width=8.5cm]{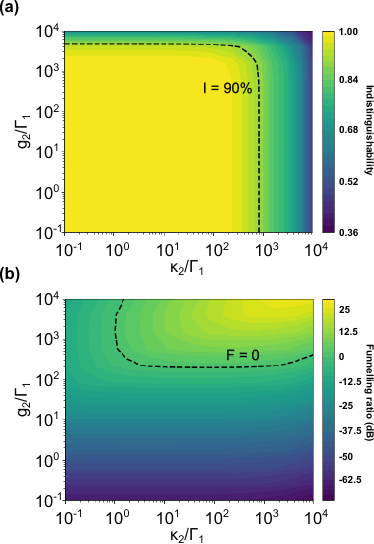}
\caption{\label{fig:I_and_F_density_gEC=0} (a) The photon indistinguishability $\left(I\right)$ expected from the hybrid architecture as a function of the outer cavity decay rate $\left(\kappa_2\right)$ and the coupling rate between that cavity and the plasmonic device $\left(g_2\right)$. The rest of the system parameters are set to $g_1 = 10^4\cdot\Gamma_1$, $\kappa_1 = 10^5\cdot\Gamma_1$ and $g_0=0$. The area of high $I$ $\left(\geq90~\%\right)$ is highlighted with a black dashed line. (b) The funneling ratio of that same system again as a function
of $\kappa_2$ and $g_2$. High values of $I$ can be reached at large values of $\kappa_2$ $\left(\sim6\cdot10^2\cdot\Gamma_1\right)$. The area that constitutes an advantage over simple linear filtering $\left(F\geq0~\text{dB}\right)$ is highlighted with a black dashed line.}
\end{figure}

Since the Hamiltonian of the system is time invariant, and assuming the Born-Markov and the rotating wave approximations \cite{10.1063/1.5115323}, the Kadanoff-Baym equations dictate that solving the underlying optical master equation and Dyson's equation allows the estimation of the time correlator $\langle b^\dagger\left(t+\tau\right)b\left(t\right)\rangle$ \cite{PhysRevLett.114.193601,haug2008quantum} and consequently that of $I$, $\beta$ and $F$ via Eq.~(\ref{eq:I_definition}), (\ref{eq:beta_definition}) and (\ref{eq:funnel_ratio_definition}) (see Supplemental Material). In our analysis, the system parameters were set to $2\gamma^*=10^4\cdot\Gamma_1$, $g_1=10^4\cdot\Gamma_1$ and $\kappa_1=10^5\cdot\Gamma_1$, with those values being based on finite-difference time-domain (FDTD) simulations of an example metallic bowtie design (see Supplemental Material). Furthermore, it has been assumed that the optical transition of the emitter coincides with the spectral center of both the plasmonic and the F-P resonator and that, at first, $g_0=0$. Solving numerically the master equation and Dyson's equation for such a system, we were able to compute the expected values of $I$ and $F$ for various combinations of the outer cavity decay rate $\kappa_2$ and the coupling rate $g_2$ between that cavity and the plasmonic device. Those results are shown in Fig.~\ref{fig:I_and_F_density_gEC=0}.    

It is clear that in such a geometry high levels of indistinguishability $(I>90\%)$ can be reached at very high values of the outer cavity decay rate. For reference, when $\kappa_2=6\cdot10^2\cdot\Gamma_1$ and $g_2 = 1.3\cdot10^3\cdot\Gamma_1$, the expected indistinguishability rises up to $I=91.69\%$, whereas the photon extraction efficiency becomes $\beta=0.465~\%$. The latter corresponds to a funneling ratio of $F=16.3$~dB which constitutes a factor of $\sim43$ times more collected photons in this system compared to a simple filter. Since the physical dimensions of the plasmonic particles are huge compared to that of the emitter, reaching values of $g_2\sim10^3\cdot\Gamma_1$ should be feasible. Furthermore, this operational regime of high $\kappa_2$ allows the generation of indistinguishable photons from such emitters via dielectric cavities of quality factors that are $\sim2$ orders of magnitude lower than those of a cascaded cavity system \cite{PhysRevLett.122.183602} and $\sim3$ orders of magnitude lower than those of a single cavity setup \cite{PhysRevLett.114.193601}. For context, a quantum emitter centered around 625~nm with a lifetime of $T_1\approx2.5$~ns, such as a typical room temperature hexagonal boron nitride (hBN) defect-based photon source \cite{tran2016quantum,Boll:20}, would achieve $I>90~\%$ with an external cavity quality factor of $Q=12558$. For a slightly shorter lifetime of $T_1\approx2$~ns, that value would decrease even more to $Q=10047$. Hence, our scheme significantly improves the practical feasibility of cavity funneling, bringing it within reach of the current experimental capabilities.

It is important to note that despite the very high levels of $g_1$ and $\kappa_1$ that have been set here, such a plasmonic resonator alone, without the outer dielectric cavity, would not be sufficient to reach the high indistinguishability regime. The relatively low ratio between $g_1$ and $\kappa_1$ in such devices limits the achievable Purcell enhancement, forbidding any significant increase in $I$. This can also be shown quantitatively for a system consisting of a single plasmonic resonator around the emitter with $g_1=10^4\cdot\Gamma_1$, $\kappa_1=10^5\cdot\Gamma_1$ and $g_2 = 0$. Then, our numerical solution, which is valid across the whole $g_1-\kappa_1$ plane, estimates a resulting indistinguishability of $I=22.19~\%$.
  
More comprehensive insight of the system's behavior can be derived via a semi-analytical approach (see Supplemental Material). Using the large difference between the plasmonic structure decay rate and the coupling rates between the emitter and the two resonators $\left(\kappa_1\gg g_1,g_2\right)$, simplified analytic expressions can be generated for the system populations and decay rates. Through those, three main photon exchange mechanisms can be identified; the first occurs between the emitter and the plasmonic resonator at an exchange rate $R_1$. The second, which is characterized by the exchange rate $R_2$, arises between that resonator and the outer dielectric cavity and the third corresponds to a direct exchange channel between the emitter and the outer cavity, with rate denoted as $R_3$. The defining expressions of these three exchange rates are 
\begin{equation} \label{eq:exchange_rates_1}
R_1 = \frac{4 g^2_1}{\Gamma_1+2\gamma^*+\kappa_1}~,
\end{equation}
\begin{equation}\label{eq:exchange_rates_2}
R_2 = \frac{4 g^2_2}{\kappa_1+\kappa_2+\phi}~,
\end{equation}
\begin{equation}\label{eq:exchange_rates_3}
R_3 = \frac{4 g^2_0}{\Gamma_1+2\gamma^*+\kappa_2}~,
\end{equation}
where $\phi={4 g^2_1}/({\Gamma_1+2\gamma^*+\kappa_2})$. Once more, the presence of the additional direct exchange rate $R_3$ that is included here differs from the expression presented by Choi et al. for a cascaded cavity system \cite{PhysRevLett.122.183602}, due to the nested geometry of our approach. 

The plasmonic implementation of the internal resonator allows for an additional conceptual simplification (see Supplemental Material). In particular, since the decay of the population of the plasmonic resonator $\left(\rho_{pp}\right)$ occurs at a rate that is orders of magnitude faster than those of the emitter and the outer cavity, the initial set of three-level rate equations can transform to a new group of two-level equations that only include the populations of the emitter $\left(\rho_{ee}\right)$ and the outer cavity $\left(\rho_{cc}\right)$. Moreover, those expressions are almost identical to those that correspond to the single emitter - single cavity funneling system that was originally presented by Grange et al. \cite{PhysRevLett.114.193601} except that in our case the decay rates $\Gamma_1$ and $\kappa$ and the exchange rate $R$ of the single cavity setup in those equations are replaced by their effective equivalents $\Gamma'_1$, $\kappa'$ and $R'$ which are defined as 
\begin{equation} \label{eq:eff_Gamma1}
\Gamma'_1 = \Gamma_1 + \frac{\kappa_1 R_1}{\kappa_1+R_1+R_2}~,
\end{equation}
\begin{equation}\label{eq:eff_kappa}
\kappa' = \kappa_2+\frac{\kappa_1 R_2}{\kappa_1+R_1+R_2}~,
\end{equation}
\begin{equation}\label{eq:eff_R}
R' = \frac{R_1 R_2}{\kappa_1+R_1+R_2} + R_3~.
\end{equation}
The equivalence between our system and the effective single-cavity model is also reflected in the corresponding Dyson equation solutions. The only difference between the resulting expressions is a minor correction introduced by the direct exchange channel between the emitter and the outer cavity when $g_0\neq 0$ (see Supplemental Material). 

\begin{figure}[t]
\centering
\includegraphics[width=8.5cm]{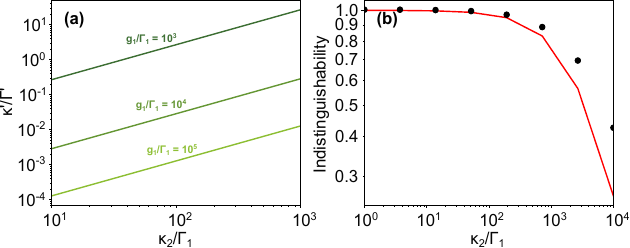}
\caption{\label{fig:eff_emitter}(a) The effective system parameter ratio $\kappa'/\Gamma'_1$ in relation to the actual ratio $\kappa_2/\Gamma_1$. Three values of $g_1$ were selected, each of which is depicted in a different shade of green. The rest of the system parameters were set at $g_2=10^2\cdot\Gamma_1$ and $\kappa_1=10^5\cdot\Gamma_1$. When $g_1\geq10^4\cdot\Gamma_1$, $\kappa'/\Gamma'_1$ becomes orders of magnitude smaller than the decay rate of the outer cavity normalised to the bare emitter decay rate $\kappa_2/\Gamma_1$. (b) The expected values of $I$ are calculated via Eq.~(\ref{eq:Indy_effective}) (continuous red curve) and via the numerical method (black dotted curve) at multiple values of $\kappa_2$ and a fixed value of $g_1 = 10^4\cdot\Gamma_1$. A good level of overlap can be observed between the two methods.}
\end{figure}

Based on the above, it can be concluded that when $g_2$ is kept at adequately low values, our system operates as a combination of two separate oscillators. The first is an effective quantum emitter whose decay rate is $\Gamma'_1$ consisting of the original photon source and the internal plasmonic antenna. The second is a new effective cavity whose decay rate is $\kappa'$, which is coupled to the effective emitter with a photon exchange rate equal to $R'$. Even though $\kappa'$ is similar to the original decay rate of the outer cavity $\left(\kappa_2\right)$, the new emitter decay rate $\Gamma'_1$ is much larger than the initial value of $\Gamma_1$ mainly due to the Purcell enhancement that is manifested by the plasmonic particles. Thus the resulting $\kappa'/\Gamma'_1$ ratio becomes much lower in this case even for high values of $\kappa_2$. This is also apparent in Fig~\ref{fig:eff_emitter}(a) that shows the evolution of the effective ratio $\kappa'/\Gamma'_1$ versus the actual ratio $\kappa_2/\Gamma_1$ for 3 values of $g_1$. It is clear that when $g_1\geq10^4\cdot\Gamma_1$, a value which is within the realistic range of plasmonic devices, the system can reach the high indistinguishability regime of $\kappa'/\Gamma'_1\leq0.1$ even when the decay rate of the outer cavity becomes as high as $\kappa_2=3\cdot10^2\cdot\Gamma_1$. Moreover, Eq.~(\ref{eq:eff_R}) dictates that the effective photon exchange rate $R'$ also remains low due to the dominating presence of $\kappa_1$ in its denominator. Thus, our system behaves almost identically to a single cavity setup that is set to operate within its $\text{low }\kappa - \text{low }g$ funnelling regime and produce high values of indistinguishability. However, in this implementation the required minimum value of the outer cavity’s quality factor is three orders of magnitude lower than in the benchmark single-cavity system.

The effective emitter model can also generate some approximate analytical expressions for the expected values of $I$ and $\beta$ in our system. These equations can be obtained from the equivalent formulas of the single cavity system simply by adapting them with the previously defined effective system parameters $\left(\Gamma'_1\text{, } \kappa'\text{ and }R'\right)$ from which we find
\begin{equation} \label{eq:Indy_effective}
I = \frac{\Gamma'_1+\frac{\kappa'R'}{\kappa'+R'}}{\left(\Gamma'_1+\kappa'+2R'\right)\left(\frac{\kappa_2+R_2+r}{\kappa'+R'}\right)}~,
\end{equation}
\begin{equation}\label{eq:beta_effective}
\beta = \frac{\kappa' R'}{\Gamma'_1 + \kappa' + 2R'}~.
\end{equation}
The validity of these approximations is verified in Fig.~\ref{fig:eff_emitter}(b) which contains the values of $I$ that correspond to multiple different levels of $\kappa_2/\Gamma_1$ using both the previously described numerical method (black dotted curve on the plot) and the analytic expressions of Eq.~(\ref{eq:Indy_effective}) (continuous red curve). The graph is presented in a double logarithmic scale and we have set $g_1=10^4\cdot\Gamma_1$, $\kappa_1=10^5\cdot\Gamma_1$, $g_2=10^2\cdot\Gamma_1$ and $g_0=0$. It is clear that within the regime of low $\kappa_2$ and low $g_2$, the values of $I$ that are estimated via the exact numerical solution are well approximated by the effective emitter model. 

\begin{figure}[t]
\centering
\includegraphics[width=8.5cm]{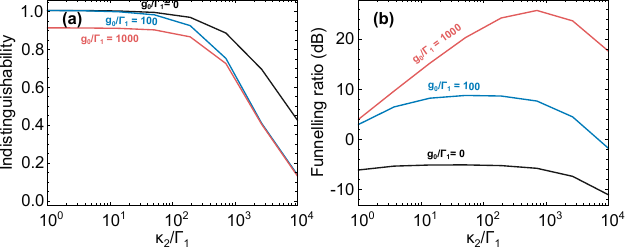}
\caption{\label{fig:g0_nonZero_linear} (a) Numerical estimation of $I$ at multiple values of $\kappa_2$ and three different levels of $g_0$ each of which is presented in a different color. (b) Same as in (a) but for the system funneling ratio, $F$.}
\end{figure}

As mentioned above, an additional difference between our approach and previous relevant attempts to improve the indistinguishability of photons generated by dephased quantum emitters is the potential for an additional direct photon exchange channel between the quantum emitter and the outer cavity, which happens at a rate $R_3$. If such a channel is absent then $g_0=0$, forcing $r$ in Eq.~(\ref{eq:Indy_effective}) to become zero as well. Then, $\kappa_2+R_2\approx\kappa'+R'$ in that same equation and the term $({\kappa_2+R_2+r})/({\kappa'+R'})$ can be neglected. Therefore, in that case both Eq.~(\ref{eq:Indy_effective}) and (\ref{eq:beta_effective}) become identical to those corresponding to the original single cavity system \cite{PhysRevLett.114.193601}. Conversely, when $g_0>0$, $R_3$ becomes the dominant term in Eq.~(\ref{eq:eff_R}). In this regime, $\kappa_2+R_2+r\approx\kappa' + R_3$ as well, forcing Eq.~(\ref{eq:Indy_effective}) and (\ref{eq:beta_effective}) to transform to 
\begin{equation} \label{eq:Indy_effective_g0Not}
I = \frac{\Gamma'_1+\frac{\kappa'R_3}{\kappa'+R_3}}{\Gamma'_1+\kappa'+2 R_3}~,
\end{equation}
\begin{equation}\label{eq:beta_effective_g0Not}
\beta = \frac{\kappa' R_3}{\Gamma'_1 + \kappa' + 2 R_3}~.
\end{equation}
From these expressions it can be seen that even though $I$ decreases as $g_0$ increases and $R_3$ becomes larger, the resulting photon efficiency $\beta$ is improved since $R_3$ appears in both the nominator and the denominator of Eq.~(\ref{eq:beta_effective_g0Not}) as well. Furthermore, since $\Gamma'_1$ still remains the dominant term in the denominator of $I$, the expected degradation rate of $I$ is much slower compared to the improvement of $\beta$. Consequently, leveraging an additional
photon exchange channel directly between the outer cavity and the quantum emitter could achieve a
significantly improved trade-off between its output photon efficiency, hence also the system funneling ratio $F$, and the photon indistinguishability. This is also intuitively reasonable as in such a scenario there is a compromise between the decreased value of $I$ due to the limited direct coupling to the low-$\kappa_2$ outer cavity and the increased value of $\beta$ that emerges from partially bypassing the lossy plasmonic resonator.

This hypothesis is verified numerically by calculating the variation of $I$ and $F$ with $\kappa_2$ for three different levels of $g_0$, with the results illustrated in Fig.~\ref{fig:g0_nonZero_linear}. These also include the case of $g_0=0$ where any direct coupling is absent and, in accordance with the settings used in the previous analyses, the rest of the system's parameters were set as follows: $g_1=10^4\cdot\Gamma_1$, $\kappa_1=10^5\cdot\Gamma_1$ and $g_2 = 10^2\cdot\Gamma_1$. It is clear that establishing a direct coupling rate of $g_0=10^2\cdot\Gamma_1$ not only maintains the photon indistinguishability at levels of 90~\% or more, but also increases the resulting funneling ratio, hence the relevant $\beta$ values as well. For instance, when $\kappa_2=10^2\cdot\Gamma_1$, $\beta$ is increased by a factor of $\sim25$ times $\left(14~\text{dB increase}\right)$ reaching $\beta=0.07\%$. Pushing $g_0$ to even higher values up to $10^3\cdot\Gamma_1$, the additional advantage can rise to 582 times $\left(27.7~\text{dB increase}\right)$ larger photon extraction efficiencies reaching the level of $\beta=1.97~\%$, without any serious compromise in the generated indistinguishability. Nevertheless, achieving such a large value of $g_{0}$ poses substantial experimental challenges. For this reason, we adopt $g_{0} = 50\cdot\Gamma_{1}$ as a practical upper bound in the present analysis. Assuming a cavity decay rate of $\kappa_2=6\cdot10^2\cdot\Gamma_1$, such a value of $g_0$ corresponds to a Purcell factor of 16.66 on the emitter's emission rate due to its interaction with the cavity.

\begin{figure}[t]
\centering
\includegraphics[width=8.5cm]{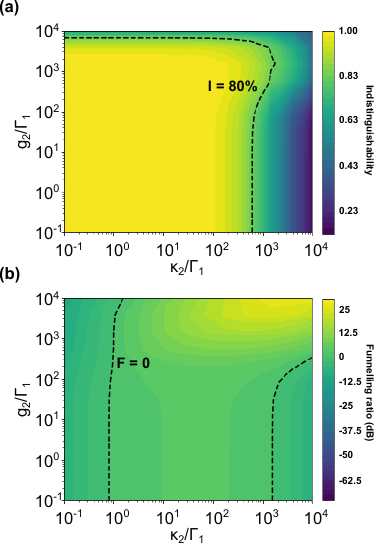}
\caption{\label{fig:g0_nonZero_density} (a) The expected values of $I$ as a function $\kappa_2$ and $g_2$ when $g_0=50\cdot\Gamma_1$. The rest of
the system parameters are set to $g_1 = 10^4\cdot\Gamma_1$ and $\kappa_1 = 10^5\cdot\Gamma_1$. (b) The funneling ratio of that system as a function of $\kappa_2$ and $g_2$. The area where this system is advantageous over linear filtering $\left(F \geq 0\right)$ is much broader here compared to Fig.~\ref{fig:I_and_F_density_gEC=0}(b).}
\end{figure}

For completeness, the dependence of the indistinguishability and the funneling ratio on both the outer cavity decay rate $\kappa_2$ and the coupling rate $g_2$ between that cavity and the plasmonic device is presented for a fixed value of $g_0=50\cdot\Gamma_1$ in Fig.~\ref{fig:g0_nonZero_density}. Comparing these results with the equivalent values that occur when $g_0=0$ (Fig.~\ref{fig:I_and_F_density_gEC=0}), it is evident that increasing $g_0$ from zero to $g_0=50\cdot\Gamma_1$, accelerates the rate of decrease of indistinguishability with $\kappa_2$, especially when $g_2$ becomes low. Still though, it can remain at levels as high as $I=80\%$ even when the outer cavity decay rate rises to $\kappa_2=6\cdot10^2\cdot\Gamma_1$ and at $I=90\%$ when $\kappa_2=3\cdot10^2\cdot\Gamma_1$. Hence, in such a configuration the practical advantage of the double resonator system over the single cavity setup is preserved. Additionally, if $g_2$ rises to the threshold value of $g_2\sim3\cdot10^2\cdot\Gamma_1$, there appears a discontinuity in the evolution of $I$ and $F$ with $g_2$ that transitions both plots to values similar to the case of $g_0=0$. Namely, if the spectral and dipole overlap between the plasmonic mode and the cavity's field increases, thus $g_2$ becomes large forcing $R_2$ to become large as well, any increase in $g_0$ becomes insignificant. On the contrary, if $g_2\leq3\cdot10^2\cdot\Gamma_1$ the funneling ratio $F$ and consequently $\beta$ become much larger when $g_0$ becomes nonzero. This is clearly apparent in Fig.~\ref{fig:g0_nonZero_density}(b) where a much larger area corresponds to system configurations where an advantage over simple linear filtering is expected $\left(F>0\right)$, even within the region of $g_2<10^2\cdot\Gamma_1$. For instance, when $g_2=70\cdot\Gamma_1$ and $\kappa_2=6\cdot10^2\cdot\Gamma_1$, not only $I>80~\%$ and $F>0$ but also $\beta$ becomes higher by a factor of $\times12$ times $\left(10.81~\text{dB increase}\right)$. Finally, it can be noticed that when $g_0=50\cdot\Gamma_1$ and $g_2<10^2\cdot\Gamma_1$ the dependence of $F$ on $g_2$ becomes negligible. This feature, which also arises directly from Eq.~(\ref{eq:beta_effective_g0Not}), allows for increased tolerance to larger mode volumes in the outer cavity and could potentially be exploited to enhance its $Q$-factor by operating on higher order modes with substantially larger physical lengths.    

In conclusion, a new approach on the cavity funneling concept has been presented here towards the efficient emission of indistinguishable photons from highly dephased solid state sources. It was shown that when a plasmonic nanoresonator is placed inside a dielectric Fabry-Perot cavity, the plasmonic structure and the emitter form a new structural element that behaves as a new effective emitter. This effective emitter exhibits a much larger decay rate, arising from both the intrinsic losses of the metallic plasmonic device, which broaden the resonant spectral width, and the Purcell enhancement induced by the plasmonic particles. As a result, external cavities containing a decay rate of up to $\kappa_2\sim10^2\cdot\Gamma_1$ can generate a deterministic stream of photons exhibiting $I\geq90~\%$ despite the high dephasing rate of the initial emitter and the low quality factors that arise. Such a high level upper bound of $\kappa_2$ forms an advantage of $\sim2$ orders of magnitude over the current state of the art value presented for a cascaded dielectric cavity \cite{PhysRevLett.122.183602} and falls within cavity quality factors already achievable in the visible spectrum.

Furthermore, and contrary to the cascaded system, the nested geometry of our system could also enables a partial direct photon exchange between the emitter and the outer cavity $\left(g_0>0\right)$ that bypasses the plasmonic resonator. We have shown that when $g_0=50\cdot\Gamma_1$ the coexistence of this mechanism along with the normal channel that funnels photons into the outer cavity via the plasmonic device could maintain an advantage over linear filtering for much lower values of $g_2$. Even, when $g_2\leq10^2\cdot\Gamma_1$, high levels of $I$ can be reached while $\times12$ times more photons per excitation cycle $\left(F\text{ and }\beta\text{ increased by }10.81~\text{dB}\right)$ can be coupled in the cavity output mode. As a result, there is a large area in the $g_2-\kappa_2$ plane that offers an advantage regarding the photon extraction efficiency over the case of simple linear filtering $\left(F>0\right)$, while also enabling significant enhancement of the system output photon efficiency leading to better funneling trade-offs. Finally, in this arrangement the system funneling ratio $F$, and hence its output photon extraction efficiency $\left(\beta\right)$, become less dependent on the size of the outer cavity resonant mode volume. This could potentially be exploited for additional enhancement of the outer cavity quality factor through operation on higher order longitudinal modes of longer cavities without severely degrading the resulting values of $\beta$.

\section*{SUPPLEMENTAL MATERIAL}

\appendix

\section{\label{sec:Estimation_of_I_and_beta}Estimation of $I$ and $\beta$}
It is assumed that initially the emitter appears in its excited state and that no photons exist in the system. Supposing then that both the rotating wave and Born-Markov approximations are valid \cite{10.1063/1.5115323}, the system only implicates the following states: $\ket{g,0,0}$ where the emitter is in its ground state with no photons being present, $\ket{e,0,0}$ which is same as before but with the emitter in its excited state, $\ket{g,1,0}$ where the emitter is in its ground state and one photon appears in the plasmonic structure and $\ket{g,0,1}$ where the emitter is again in its ground state and one photon appears in the dielectric cavity mode. Then, if the reduced Planck’s constant is equal to 1 and there is no detuning between the emitter and any of the optical resonators, the Hamiltonian of this system can be expressed as
\begin{equation} \label{eq:Hamiltonian}
H = g_1\left(a^\dagger\sigma^-+\sigma^+a\right)+g_2\left(a^\dagger b + b^\dagger a\right) + g_0\left(b^\dagger\sigma^- + \sigma^+ b\right)\,,
\end{equation}
where $a$, $a^\dagger$ are the bosonic creation and annihilation operators of the plasmonic mode, $b$, $b^\dagger$ those of the dielectric cavity's mode and $\sigma^+=\ket{e}\bra{g}$ and $\sigma^-=\ket{g}\bra{e}$ the emitter electronic energy level raising and lowering operators. Since no coherent coupling is expected between $\ket{g,0,0}$ and any of the other 3 states the only subset of interest within the total Hilbert space is the subspace that is spanned by the states $\ket{e,0,0}$, $\ket{g,1,0}$ and $\ket{g,0,1}$. 

The corresponding optical master equation that describes the system's density operator $\rho\left(t\right)$ is the following
\begin{equation} \label{eq:master}
\frac{d\rho}{dt} = -i\left[H,\rho\right]+L_{qe}\left[\rho\right]+L_{deph}\left[\rho\right]+L_{p}\left[\rho\right]+L_{c}\left[\rho\right]\,.
\end{equation}
The $L_i\left[\rho\right]$ terms are the relevant Lindblad contributions that express all the dephasing effects which occur due to the emitter's spontaneous emission $\left(L_{qe}\left[\rho\right]\right)$, its pure dephasing mechanisms $\left(L_{deph}\left[\rho\right]\right)$, the outer cavity's decay $\left(L_{c}\left[\rho\right]\right)$ and the plasmonic structure's internal losses $\left(L_{p}\left[\rho\right]\right)$. The matrix form of those terms in the $\{\ket{e,0,0},\ket{g,1,0},\ket{g,0,1}\}$ basis is
\begin{equation} \label{eq:LQe_1-2}
L_{qe}\left[\rho\right] = -\Gamma_1 \begin{bmatrix}
\rho_{ee} & \frac{\rho_{ep}}{2} & \frac{\rho_{ec}}{2}\\
\frac{\rho_{pe}}{2} & 0 & 0\\
\frac{\rho_{ce}}{2} & 0 & 0
\end{bmatrix} \,,
\end{equation}
\begin{equation} \label{eq:Ldeph_1-2}
L_{deph}\left[\rho\right] = -\gamma^* \begin{bmatrix}
0 & \rho_{ep} & \rho_{ec}\\
\rho_{pe} & 0 & 0\\
\rho_{ce} & 0 & 0
\end{bmatrix}\,,
\end{equation}
\begin{equation} \label{eq:Lp_1-2}
L_{p}\left[\rho\right] = -\kappa_1 \begin{bmatrix}
0 & \frac{\rho_{ep}}{2} & 0\\
\frac{\rho_{pe}}{2} & \rho_{pp} & \frac{\rho_{pc}}{2}\\
0 & \frac{\rho_{cp}}{2} & 0
\end{bmatrix}\,.
\end{equation}
\begin{equation} \label{eq:Lc_1-2}
L_{c}\left[\rho\right] = -\kappa_2 \begin{bmatrix}
0 & 0 & \frac{\rho_{ec}}{2}\\
0 & 0 & \frac{\rho_{pc}}{2}\\
\frac{\rho_{ce}}{2} & \frac{\rho_{cp}}{2} & \rho_{cc}
\end{bmatrix}\,,
\end{equation}
whereas the elements of these matrices are defined from the expressions below $\rho_{ee}=\bra{e,0,0}\rho\ket{e,0,0}$, $\rho_{pp}=\bra{g,1,0}\rho\ket{g,1,0}$, $\rho_{cc}=\bra{g,0,1}\rho\ket{g,0,1}$, $\rho_{ep}=\bra{e,0,0}\rho\ket{g,1,0}$, $\rho_{ec}=\bra{e,0,0}\rho\ket{g,0,1}$, $\rho_{pc} = \bra{g,1,0}\rho\ket{g,0,1}$ and their complex conjugates.

Based on their definitions $I$, $\beta$ and $F$ can be estimated if the quantity $\langle b^\dagger\left(t+\tau\right)b\left(t\right)\rangle$ is known. Using the Kadanoff-Baym equations \cite{PhysRevLett.114.193601,PhysRevLett.122.183602,haug2008quantum}, this quantity can be estimated as follows
\begin{align} \label{eq:time_correlation}
    \langle b^\dagger\left(t+\tau\right)b\left(t\right)\rangle = G^{r}_{ce}\left(\tau\right) \rho_{ec}\left(t\right) + G^{r}_{cp}\left(\tau\right) \rho_{pc}\left(t\right) +\nonumber\\ G^{r}_{cc}\left(\tau\right) \rho_{cc}\left(t\right) \,,
\end{align}
with $G_{ij}^r\left(\tau\right)$ the matrix elements of the system's retarded Green's function whose Fourier transform can be estimated via Dyson's equation 
\begin{equation} \label{eq:Dyson}
    G^r\left(\omega\right) = \left(\omega - H - \Sigma^r\left(\omega\right) \right)^{-1} \,.
\end{equation}
In the selected basis, the self energy term in the above equation is defined as
\begin{equation} \label{eq:Self_1-2}
     \Sigma^r\left(\omega\right) =  -i \begin{bmatrix}
\frac{\Gamma_1}{2}+\gamma^* & 0 & 0\\
0 & \frac{\kappa_1}{2} & 0\\
0 & 0 & \frac{\kappa_2}{2}
\end{bmatrix} \,.
\end{equation}
Solving numerically Eq.~(\ref{eq:master}) to get $\rho\left(t\right)$ and Eq.~(\ref{eq:Dyson}) to get $G^r\left(\tau\right)$ allows the computation of $\langle b^\dagger\left(t+\tau\right)b\left(t\right)\rangle$ via Eq.~(\ref{eq:time_correlation}). Subsequently, inserting $\langle b^\dagger\left(t+\tau\right)b\left(t\right)\rangle$ into Eq.~(\ref{eq:I_definition_supp}) and Eq.~(\ref{eq:beta_definition_supp}) enables the evaluation of $I$ and $\beta$ for any combination of the system's constructive parameters $\left(g_0, g_1, g_2, \kappa_1, \kappa_2\right)$ 
\begin{equation}
I = \frac{ \int_{0}^\infty \int_{0}^\infty |\langle b^\dagger\left(t+\tau\right) b\left(t\right)\rangle|^2 d\tau dt}{\int_{0}^\infty \int_{0}^\infty \langle b^\dagger\left(t\right) b\left(t\right)\rangle\langle b^\dagger\left(t+\tau\right) b\left(t+\tau\right)\rangle d\tau dt} \;,
\label{eq:I_definition_supp}
\end{equation}
\begin{equation}
    \beta = \kappa_2 \int_{0}^\infty \langle b^\dagger\left(t\right) b\left(t\right) \rangle dt \;,
\label{eq:beta_definition_supp}
\end{equation}
\begin{equation}
    F = 10 \log\left(\frac{2\gamma^*}{\Gamma_1}\beta I\right)
    \label{eq:funnel_ratio_definition_supp} \,.
\end{equation}
Then, using the estimated values of $I$ and $\beta$ in Eq.~(\ref{eq:funnel_ratio_definition_supp}) generates the value of the funneling ratio $F$ as well

\section{\label{sec:Analytical_approx}Analytic approximation}
It is possible to obtain analytic approximations for the populations of any system of nested optical resonators. The exact approximations that generate those solutions, however, depend heavily on the relative size of the system parameters, with each regime of operation yielding a different set of simplifications. Here, contrary to the purely dielectric nature of the cascaded system suggested by Choi et al. \cite{PhysRevLett.122.183602}, it is assumed that the first structure is a plasmonic nanoresonator. Such a structure is expected to have much smaller mode volumes, hence high values of $g_1$, and a very large decay rate $\kappa_1$ due to its high losses. Therefore, within the regime of weak coupling between that resonator and the outer cavity, it is safe to assume that $2g_0\ll\Gamma_1+2\gamma^*+\kappa_2$, $2g_1\ll\Gamma_1+2\gamma^*~+\kappa_1$ and $2g_2\ll\kappa_1+\kappa_2$, which allow the following adiabatic eliminations $\frac{d\rho_{ec}}{dt},\frac{d\rho_{ce}}{dt},\frac{d\rho_{ep}}{dt}, \frac{d\rho_{pe}}{dt}, \frac{d\rho_{pc}}{dt}\text{ and }\frac{d\rho_{cp}}{dt}\sim0$. Utilizing these eliminations along with the general dominance of $\kappa_1$ over all the other parameters, Eq.~(\ref{eq:master}) results in the following expressions for the off-diagonal elements of the system's density operator
\begin{align} \label{eq:r_ec}
\rho_{ec} = \frac{2i g_0\left(\rho_{ee}-\rho_{cc}\right)}{\Gamma_1+2\gamma^*+\kappa_2}-\frac{4g_1g_2\left(\rho_{ee}-\rho_{pp}\right)}{\left(\Gamma_1+2\gamma^*+\kappa_2\right)\left(\Gamma_1+2\gamma^*+\kappa_1\right)}\nonumber\\
+\frac{4g_1g_2\left(\rho_{pp}-\rho_{cc}\right)}{\left(\Gamma_1+2\gamma^*+\kappa_2\right)\left(\kappa_1+\kappa_2+\phi\right)}\,,
\end{align}
\begin{align}\label{eq:r_ce}
\rho_{ce} = \frac{2ig_0\left(\rho_{cc}-\rho_{ee}\right)}{\Gamma_1+2\gamma^*+\kappa_2}+\frac{4g_2g_1\left(\rho_{pp}-\rho_{ee}\right)}{\left(\Gamma_1+2\gamma^*+\kappa_2\right)\left(\Gamma_1+2\gamma^*+\kappa_1\right)}\nonumber\\-\frac{4g_2g_1\left(\rho_{cc}-\rho_{pp}\right)}{\left(\Gamma_1+2\gamma^*+\kappa_2\right)\left(\kappa_2+\kappa_1+\phi\right)}\,,
\end{align}
\begin{equation}\label{eq:r_ep}
\rho_{ep} = \frac{2ig_1\left(\rho_{ee}-\rho_{pp}\right)}{\Gamma_1+2\gamma^*+\kappa_1}\,,
\end{equation}
\begin{equation}\label{eq:r_pe}
\rho_{pe} = \frac{2ig_1\left(\rho_{pp}-\rho_{ee}\right)}{\Gamma_1+2\gamma^*+\kappa_1}\,,
\end{equation}
\begin{align}\label{eq:r_pc}
\rho_{pc} = \frac{2ig_2\left(\rho_{pp}-\rho_{cc}\right)}{\kappa_1+\kappa_2+\phi}-\frac{4g_0g_1\left(\rho_{pp}-\rho_{ee}\right)}{\left(\Gamma_1+2\gamma^*+\kappa_1\right)\left(\kappa_1+\kappa_2+\phi\right)}\,,
\end{align}
\begin{equation}\label{eq:r_cp}
\rho_{cp} = \frac{2ig_2\left(\rho_{cc}-\rho_{pp}\right)}{\kappa_1+\kappa_2+\phi}+\frac{4g_0g_1\left(\rho_{ee}-\rho_{pp}\right)}{\left(\Gamma_1+2\gamma^*+\kappa_1\right)\left(\kappa_1+\kappa_2+\phi\right)}\,,
\end{equation}
with $\phi=\frac{4g^2_1}{\Gamma_1+2\gamma^*+\kappa_2}$. Inserting these expressions into the system's master equation (Eq.~(\ref{eq:master})) and solving it for the system's populations $\left(\rho_{ee}\left(t\right),\rho_{pp}\left(t\right)\text{ and }\rho_{cc}\left(t\right)\right)$, its rate equations can be extracted in the following manner
\begin{align} \label{eq:rate}
\begin{bmatrix}
\dot\rho_{ee}\left(t\right)\\\dot\rho_{pp}\left(t\right)\\\dot\rho_{cc}\left(t\right)
\end{bmatrix} = A\cdot\begin{bmatrix}
\rho_{ee}\left(t\right)\\\rho_{pp}\left(t\right)\\\rho_{cc}\left(t\right)
\end{bmatrix}\,,
\end{align}
with matrix $A$ being defined as
\begin{align}
A = \begin{bmatrix}
-\Gamma_1-R_1-R_3 & R_1 & R_3\\
R_1 & -\kappa_1-R_1-R_2 & R_2\\
R_3 & R_2 & -\kappa_2-R_2-R_3
\end{bmatrix}\,.
\end{align}
The photon exchange rates that appear in the equation above are defined as
\begin{equation}
\begin{aligned}\label{eq:exchange_rates}
R_1 = \frac{4g_{1}^2}{\Gamma_1+2\gamma^*+\kappa_1}
\\
R_2 = \frac{4g_{2}^2}{\kappa_1+\kappa_2+\phi}
\\
R_3 = \frac{4g_{0}^2}{\Gamma_1+2\gamma^*+\kappa_2}
\end{aligned}\,.
\end{equation}
Three main photon exchange mechanisms can be identified from Eq.~(\ref{eq:rate}); one between the emitter and the plasmonic resonator which is characterized by the exchange rate $R_1$, another between that plasmonic structure and the outer dielectric cavity and which is quantified by the exchange rate $R_2$ and a third one that directly couples the emitter to the outer cavity via the exchange rate $R_3$. As is apparent, the hereby presented rate equations are almost identical to those presented in the case of the cascaded system \cite{PhysRevLett.122.183602}. The key difference between them is the presence of the third direct exchange channel between the dielectric cavity and the emitter. This channel emerges due to the nested topology of our system and could not be implemented in any relevant cascaded geometry. 

\section{Effective emitter simplification}\label{sec:effective_emitter}
The unique nature of plasmonic structures allows for an additional simplification of Eq.~(\ref{eq:rate}). The fast dissipation of its population, which stems from its very large decay rate $\kappa_1$, allows the assumption that $\frac{d\rho_{pp}}{dt}\sim0$ after a short period of time. Applying this into the rate equations above generates the following simplified system
\begin{align} \label{eq:rate_effective}
\begin{bmatrix}
\dot\rho_{ee}\left(t\right)\\\dot\rho_{cc}\left(t\right)
\end{bmatrix} = \begin{bmatrix}
-\Gamma'_1-R' & R'\\
R' & -\kappa'-R'\end{bmatrix}\cdot\begin{bmatrix}
\rho_{ee}\left(t\right)\\\rho_{cc}\left(t\right)
\end{bmatrix}\,.
\end{align}
It consists of a set of effective constructive parameters which are defined as
\begin{equation}
\begin{aligned}\label{eq:efficient_params_supp}
\Gamma'_1 = \Gamma_1 + \frac{\kappa_1 R_1}{\kappa_1+R_1+R_2}
\\
\kappa' = \kappa_2 + \frac{\kappa_1 R_2}{\kappa_1+R_1+R_2}
\\
R' = \frac{R_1 R_2}{\kappa_1+R_1+R_2} + R_3
\end{aligned}\,,    
\end{equation}
with the population of the plasmonic device being equal to
\begin{equation}\label{eq:effective_emitter_rPP}
\rho_{pp} = \frac{R_1\rho_{ee} +R_2\rho_{cc}}{\kappa_1+R_1+R_2}\,.
\end{equation}
Then, since $\kappa_1\gg R_1,R_2$ and $\kappa_2\gg R_2$, it can also be assumed from Eq.~(\ref{eq:efficient_params_supp}) that $\kappa'\approx\kappa_2$ and $\Gamma'_1\approx \Gamma_1\cdot\left(1+F^{*}_p\right)$, where $F^{*}_p=\frac{4g^{2}_1}{\left(\Gamma_1+2\gamma^*+\kappa_1\right)\Gamma_1}$ is a generalized form of the Purcell
factor that would be imposed on the emitter if the system
only contained the plasmonic device \cite{PhysRevB.81.245419,doi:10.1021/acsphotonics.7b00475}.

\begin{figure}[t]
\centering
\includegraphics[width=8.5cm]{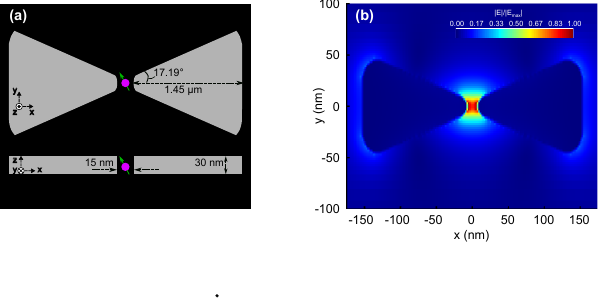}
\caption{\label{fig:bowtie_geometry}(a) The suggested design for the bowtie structure. (b) The spatial profile of the electric field at its resonant wavelength $\left(\lambda=625.13~\text{nm}\right)$.}
\end{figure}

Using the inverse Fourier transform into Eq.~(\ref{eq:Dyson}), it can be shown that Eq.~(\ref{eq:time_correlation}) is dominated by the term $G_{cc}^{r}\left(\tau\right)\rho_{cc}\left(t\right)$. Furthermore, since both $\frac{dG_{pc}^{r}}{d\tau}$ and $\frac{dG_{ec}^{r}}{d\tau}$ decay rapidly, hence they can be set to zero, it can be derived that
\begin{equation}\label{eq:Greens_approx}
G_{cc}^{r} \sim e^{-\frac{\kappa_2+R_2+r}{2}\tau}\,,
\end{equation}
where the additional term $r$ only occurs if $g_0\neq0$ and is defined as
\begin{equation} \label{eq:r_small}
r = \frac{4g_{0}^2}{\Gamma_1+2\gamma^*+R_1}\,.
\end{equation}
Then, in a similar manner to the single cavity funneling system \cite{PhysRevLett.114.193601}, our effective emitter model suggests that also in our system the indistinguishability can be given by the following expression
\begin{equation} \label{eq:Indy_effectiveEmitter_approx}
I \sim \frac{ \int_{0}^\infty |\rho_{cc}\left(t\right)|^2 dt\int_{0}^\infty |G_{cc}^r\left(\tau\right)|^2 d\tau}{\frac{1}{2}\left(\int_{0}^\infty \rho_{cc}\left(t\right) dt\right)^2} \;.
\end{equation}
Combining the above equation with Eq.~(\ref{eq:Greens_approx}), along with the general form that can be obtained for $\rho_{cc}\left(t\right)$ from Eq.~(\ref{eq:rate_effective}) and Vieta's formulas for the sum and product of its eigenvalues, it is possible to generate the following analytic expression for $I$
\begin{equation} \label{eq:indy_effective_supp}
I = \frac{\Gamma'_1+\frac{\kappa'R'}{\kappa'+R'}}{\left(\Gamma'_1+\kappa'+2R'\right)\left(\frac{\kappa_2+R_2+r}{\kappa'+R'}\right)}\,.
\end{equation}
Similarly, the photon extraction efficiency can also be estimated as
\begin{equation} \label{eq:beta_effective_supp}
\beta =  \frac{\kappa' R'}{\Gamma'_1 + \kappa' + 2R'}
\end{equation}

\section{Example design of a bowtie resonator}\label{sec:bowtie}

\begin{figure}[t]
\centering
\includegraphics[width=8.5cm]{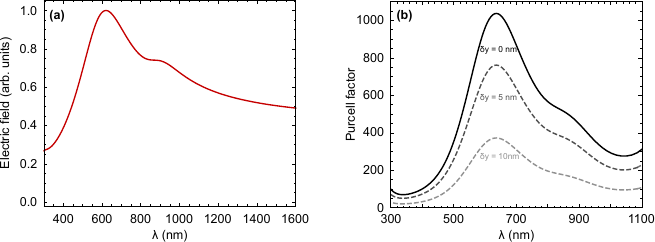}
\caption{\label{fig:bowtie_spectrum_purcell}(a) The spectral distribution of the bowtie’s resonant mode as a function of the excitation wavelength. It corresponds to an estimated $\text{FWHM}\approx451.949$~THz. (b) The calculated spectrum of the Purcell factor for 3 different levels of deviation between the emitter's position and the center of the bowtie. All these displacements are considered to occur along the device's vertical axis (y).}
\end{figure}

For completeness, an example design of a suitable bowtie resonator has been included as well. Since hexagonal boron nitride (hBN) constitutes one of the most popular options regarding room temperature solid state emitters, an hBN source has been assumed as the emitter of choice, with a target emission wavelength near 625~nm. Since the absorption spectrum of gold (Au) can be problematic near such emission lines, aluminum (Al) has been selected instead as the fabrication material of the suggested structure \cite{doi:10.1021/jp058091f,BLABER2007184}. Using finite-difference time-domain (FDTD) simulations, the tip angle and the length of the plasmonic particles were optimized in order to maximise the electric field resonance at the selected wavelength in the center of the structure. The resulting geometry is presented in Fig.~\ref{fig:bowtie_geometry}(a), while a heat map of the spatial distribution of its resonant electric field is illustrated in Fig.~\ref{fig:bowtie_geometry}(b). 

The spectral profile of that mode is presented in Fig.~\ref{fig:bowtie_spectrum_purcell}(a). As expected, the bowtie resonant wavelength is centered at 625.13~nm and the plasmonic cavity decay rate that corresponds to the full width at half maximum (FWHM) of this mode is equal to $\kappa_1=2839.68\cdot10^{12}$~rad $s^{-1}$. Assuming an emitter with a lifetime of $T_1=2.5$~ns, this results in a ratio of $\kappa_1 = 7.099\cdot10^6\cdot\Gamma_1$.

\begin{figure}[t]
\centering
\includegraphics[width=8.5cm]{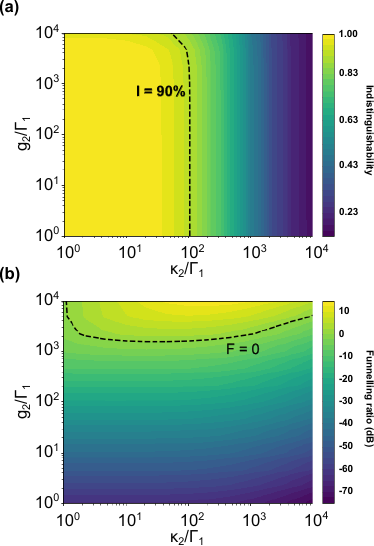}
\caption{\label{fig:bowtie_densities}(a) The photon indistinguishability that is estimated for the selected values of $g_1=3.417\cdot10^4\cdot\Gamma_1$ and $\kappa_1 = 7.099\cdot10^6\cdot\Gamma_1$ given as a function of $g_2$ and $\kappa_2$. These levels of $g_1$ and $\kappa_1$ correspond to a conservative estimation of the suggested bowtie's performance. (b) Same as in (a) but for the system's funneling ratio.}
\end{figure}

The variation with wavelength of the Purcell factor that is imposed by this device on the quantum emitter is also included in Fig.~\ref{fig:bowtie_spectrum_purcell}(b), for 3 different levels of vertical deviation between the emitter's position and the center of the device. These include the detrimental case of zero misalignment $\left(\delta y=0\right)$ as well. These estimations were conducted again via FDTD by dividing the average radiated power that occurs when the plasmonic particles are present with its corresponding value when they are not \cite{PhysRevLett.96.113002,Novotny_Hecht_2012}. Even though this method cannot capture any non-classical loss effects, such as the enhancement of non-radiative processes within the emitter \cite{2c9964a85fd94b168c3b5a4596f06e6b,PhysRevLett.96.113002}, in the suggested geometry the gap between the plasmonic particles is large enough so that any such phenomena should not be significant. The calculated Purcell factor at $\lambda=625.13$~nm is $F_p = 1029$ when the emitter's position is optimized and $F_p = 369.304$ when $\delta y = 10$~nm. Since the Purcell factor can be defined as
\begin{equation}\label{eq:Purcell_factor_supp}
F_p = \frac{4 g^{2}_1}{\kappa_1\Gamma_1}\,,
\end{equation}
these values generate coupling rates equal to $g_1 = 4.273\cdot10^4\cdot\Gamma_1$ and $g_1 = 2.56\cdot10^4\cdot\Gamma_1$ correspondingly.

In any practical implementation of such a plasmonic structure, the actual Purcell enhancement that would occur would probably be much lower than the optimized value of 1029. This could be imposed by any unavoidable deviations between its position and the center of the structure due to the limited spatial resolution of lithography. The indistinguishability and funneling ratio values that occur for an intermediate value of $g_1 = 3.417\cdot10^4\cdot\Gamma_1$ and $\kappa_1 = 7.099\cdot10^6\cdot\Gamma_1$ are presented in Fig.~\ref{fig:bowtie_densities} as a function of the outer cavity's decay rate $\left(\kappa_2\right)$ and its coupling rate with the plasmonic device $\left(g_2\right)$. It has been assumed here that $g_0=0$.

\bibliography{Fasoulakis_arXiv}

\end{document}